\documentclass[a4paper]{article}
\usepackage[a4paper, margin=1.5in]{geometry}
\usepackage[utf8]{inputenc}
\usepackage{hyperref}
\usepackage{cite}
\usepackage{graphicx}
\graphicspath{{figures/}}
\usepackage{caption}
\usepackage{authblk}
\usepackage{algorithm}
\usepackage{algpseudocode}

\hypersetup{
  pdfinfo={
    Title={Science Concierge: A fast content-based recommendation system for scientific publicationse},
    Author={Achakulvisut et. al.},
    Subject={Information retrieval},
    Keywords={recommendation system, topic modeling}
  }
}
\usepackage{pdfpages}

\title{Science Concierge: A fast content-based recommendation system for scientific publications}

\author[1]{Titipat Achakulvisut\thanks{titipat.a@u.northwestern.edu}}
\author[3]{Daniel E. Acuna\thanks{This work was done while the author was at the Rehabilitation Institute of Chicago and Northwestern University}}
\author[4]{Tulakan Ruangrong}
\author[1,2]{Konrad Kording}
\affil[1]{Department of Biomedical Engineering, Northwestern University}
\affil[2]{Sensory Motor Performance Program, Rehabilitation Institute of Chicago}
\affil[3]{School of Information Studies, Syracuse University}
\affil[4]{Department of Biomedical Engineering, Mahidol University}

\begin{document}
\date{}
\maketitle

\section*{Abstract}

Finding relevant publications is important for scientists who have to cope with exponentially increasing numbers of scholarly material. Algorithms can help with this task as they help for music, movie, and product recommendations. However, we know little about the performance of these algorithms with scholarly material. Here, we develop an algorithm, and an accompanying Python library, that implements a recommendation system based on the content of articles. Design principles are to adapt to new content, provide near-real time suggestions, and be open source. We tested the library on 15K posters from the Society of Neuroscience Conference 2015.
Human curated topics are used to cross validate parameters in the algorithm and produce a similarity metric that maximally correlates with human judgments. We show that our algorithm significantly outperformed suggestions based on keywords.
The work presented here promises to make the exploration of scholarly material faster and more accurate.

\section{Introduction}

Recommendation systems are routinely used to suggest items based on customers’ past preferences. These systems have proven useful for music, movies, news, and retail in general \cite{manning2008introduction}. In contrast, to find new scientific literature, researchers rely mostly on author-provided keywords, titles, author names, and references. These procedures are likely to be biased \cite{chavalarias2010science} and also provide an
unintended rich get richer (or Matthew's) effects \cite{petersen2011quantitative}.
This problem is more pronounced during conferences where appropriate keywords may not even exist, let alone citations.
An application of recommendation systems to suggest scholarly material based on the researcher’s preferences thus promises to speed up literature search and increase relevance.

There are multiple recommendation systems that use either the personal preferences of a new user (e.g., content-based recommendations) or exploit the similarity between the new users’ preferences and previous users’ preferences (e.g., collaborative filtering). Most such system are available for commercial software, such as news \cite{li2010contextual}, movie \cite{bell2007lessons}, and music \cite{Ali11parallelcollaborative} applications. In contrast, few projects address scientific literature search.

Scientific literature is different from other applications of recommendation systems.
Scientist are very specialized and they tend to become even more specialized
as they progress in their careers. Therefore, understanding the differences in such
fine grained topics is challenging. For example, for the Society for Neuroscience
conference, there are more than 500 hand-curated areas of research (\url{github.com/titipata/science_concierge/wiki/Topic-in-SfN-2015}). The entire
library of medicine contains more than 25K hand-curated vocabultary terms to denote research areas \cite{lipscomb2000medical}. By comparison, there are around 400 genres and sub-genres of music.
Additionally, the priority discovery credit system of science makes scientists hide their discoveries until full publication.
This means that entirely new fields may appear overnight, forcing scientific discovery systems
to quickly adapt. Scientific literature is different from other fields and therefore
specific requirements need to be addressed.

There have been several attempts at producing recommendation systems for scientific literature.
In \cite{yoneya2007pure}, the authors presented a content-based recommendation system that works on PubMed datasets. In \cite{gipp2009scienstein}, the Scienstein system combined a large set of criteria for providing literature recommendation. In \cite{wang2011collaborative}, the authors presented a topic-based recommendation system based on a Latent Dirichlet Allocation (LDA) model. Finally, \cite{wesley2016babel} presents a recommendation system based on citations. It is unclear how well these use the content of the
manuscripts and how they work internally since many of them are not open sourced.

Here we introduce Science Concierge (\href{https://github.com/titipata/science_concierge}{\texttt{github.com/titipata/science\_concierge}}), an open source Python library that implements a recommendation system for literature search. Briefly, the library uses a scalable vectorization of documents through online Latent Semantic Analysis (LSA) \cite{manning2008introduction}. For the recommendation part, it pairs the Rocchio Algorithm \cite{rocchio1971relevance} with a large-scale approximate nearest neighbor search based on ball trees \cite{shakhnarovish2005indyk}. The library aims at providing responsive content-based recommendations utilizing only user’s votes. Science Concierge, then, provides an open source solution to content-based scientific discovery.

We tuned and tested the algorithm on a collection of scientific posters from the largest Neuroscience conference in the world, Society for Neuroscience (SfN) 2015. First, we cross-validated the LSA model to capture most of the variance contained in the topics. Second, we tuned the parameters of the algorithm to recommend posters that maximally resembled human curated classifications into poster sessions. We showed that our algorithm significantly outperformed a popular alternative based on keywords, improving suggestions further as it learned more from the user. A front-end interface that uses the algorithm in the back end is available at \href{http://www.scholarfy.net}{\texttt{http://www.scholarfy.net}}, where we used data from Society for Neuroscience (SfN) 2015 conference.

\section{Materials and methods} \label{methods}

\subsection{Conference dataset}

Conferences are events that need a rapid understanding of trends. The time between submissions, acceptances, and presentations is typically much shorter than in journals. This makes it crucial for recommendation systems to quickly scan the documents of the conference and let scientists discover materials fast. This is one reason why we focus on testing our system using conference data.

We obtained a license from the Society for Neuroscience (SfN) for the Neuroscience 2015 conference. This is the largest conference in Neuroscience in the world. This dataset included 14,718 posters and talks distributed over 500 sessions spanning 5 days. Not all the content of the conference had abstracts available (e.g., talks) and therefore they were dropped from the analysis. The dataset is not publicly available but an academic license may be requested from the Society (\href{https://www.sfn.org}{\texttt{http://www.sfn.org}}).

\subsection{Content-based recommendation of scientific documents}

There are three main design principles behind Science Concierge. First, it aims at using the content of the documents rather than collaborative filter to prevent the Mathew's effect, commonly found in recommendation systems  \cite{petersen2011quantitative}. This effect can be detrimental for scientific exploration. Second, it aims at providing  suggestions as fast as possible. This means that users should get feedback as soon as they vote one item as relevant or irrelevant. Finally, it aims at being validated using some external input. Below we describe the methods to achieve these three goals.

\begin{figure}[!ht]
\centering
\includegraphics[width=3.5in]{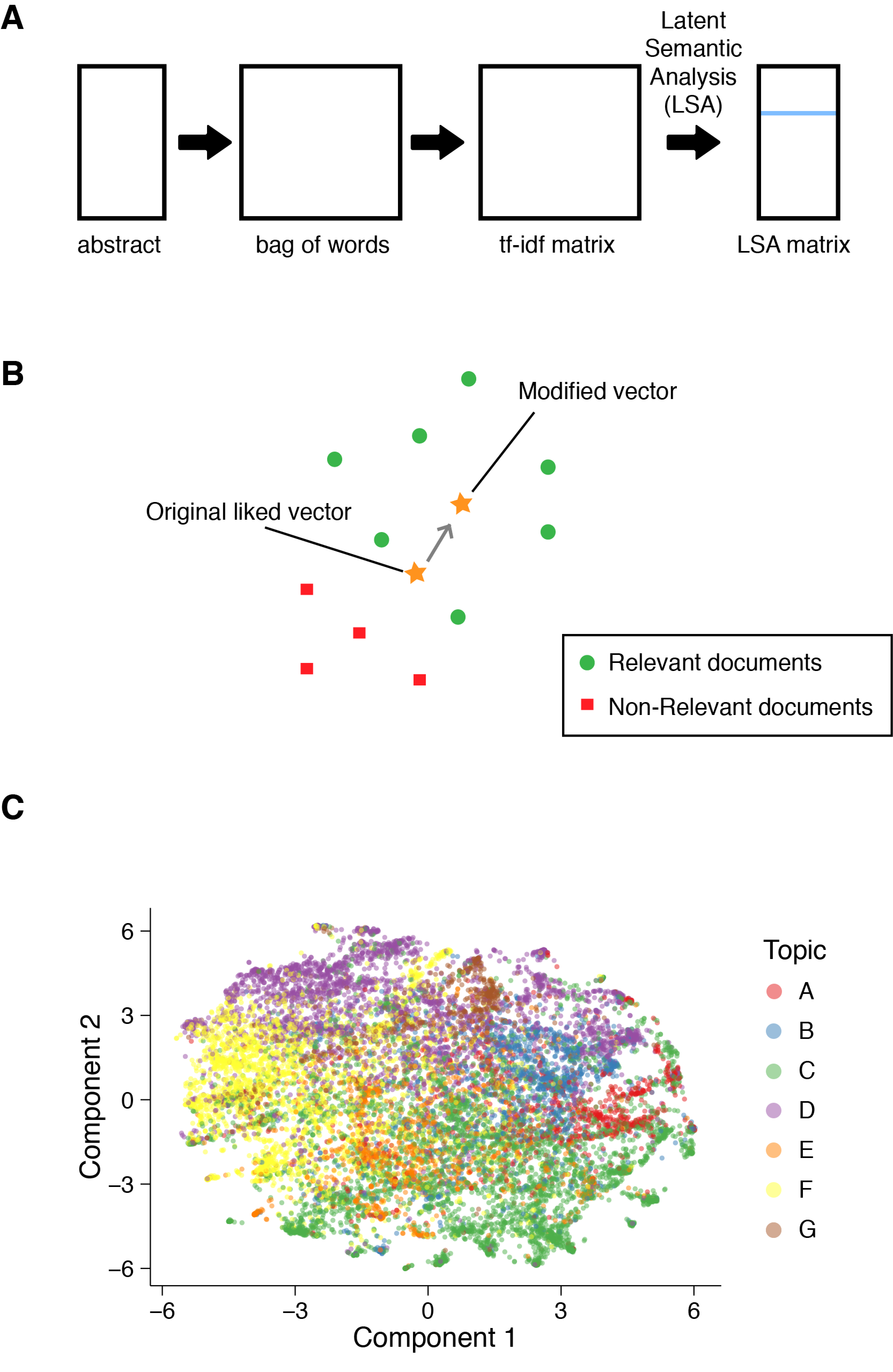}
\caption*{\textbf{Fig 1.} \textbf{Vector representation of documents} (A) Schematic of the workflow for converting abstracts into vector representations (see Algorithm 1) (B) Schematic of Rocchio Algorithm (C) Projection of SfN abstract vectors to 2-dimensional space using $t$-SNE color coded by human curated sessions from A to G}
\end{figure}

\subsection{Text preprocessing and term weighting schemes}

Documents are tokenized by removing English stop words and stemming using
the Porter Stemming algorithm \cite{porter1980algorithm}. The terms in the
documents are uni-grams and bi-grams. In the experiment, terms which appear fewer than 3 times were removed or terms that have tf-idf (explained below) greater than 0.8 were removed.

These terms can be weighted in different
manners by considering the local structure of each document as compared
to the global structure of all documents. Different weighting schemes
exist in the literature and we try some of the most popular, including
simple term frequency (term frequency),
term-frequency inverse document-frequency (tf-idf), and
log-entropy \cite{lee2005empirical, pincombe2004comparison}.
We additionally try a word vector representation that we explain
in more detail later.

Term frequency, tf-idf, and log-entropy can all be computed based on the
following term statistics. Let $f_{ij}$ be the frequency of term $i$
in document $j$. Term frequency uses the matrix $f_{ij}$ directly.
Tf-idf log transforms and reweights each term frequency by a global factor as follows

\begin{equation}
\mbox{tf-idf}_{ij} = (1 + \log f_{ij}) \times \log \frac{n}{f_i + 1}
\label{eq:tfidf}
\end{equation}
where $f_i$ is number of documents that contains term $i$ and $n$ is the total number of documents. For the log-entropy transformation, we first find the global entropy for each term $i$ as follows

\begin{equation}
g_i = 1 + \sum_j \frac{p_{ij} \log_2 p_{ij}}{\log_2 n} \mbox{ where } p_{ij} = \frac{f_{ij}}{\sum_j f_{ij}}
\label{eq:globalweight}
\end{equation}

Finally, the log-entropy of term $i$ in document $j$ or $l_{ij}$ can be written as

\begin{equation}
l_{ij} = \log_2(1 + f_{ij}) \times g_i
\label{eq:localweight}
\end{equation}

Additionally,
we use an alternative representation of a document based on word
vectors. The word vector represention \cite{mikolov2013distributed} allows
to approximate the co-ocurrence of terms, and therefore captures much better the meaning of words.
This kind of representation has been used successfully for synonym finding and translation tasks. In this article, we compute a 150 dimensional representation
of the word vectors training from the abstracts of SfN 2015 and average the words that appear in each document.

\subsection{Latent Semantic Analysis}

LSA is a simple topic modeling technique based on singular value decomposition \cite{landauer1998introduction}. The
different weighting schemes describe in the previous section
are transformed using Latent Semantic Analysis (LSA) to reduce noise and improve smoothness. This technique decomposes the weighting matrix $X$ as a product of a left singular vectors ($U$), a diagonal singular value matrix ($S$) and a right singular vectors ($V$) as $X = USV^\top$, where the diagonal singular values in $S$ are ranked from highest to lowest. The number of vectors to choose depends on how accurately we want to capture the matrix $X$ \cite{bishop2006pattern}, i.e. $X \approx U_r S_r V_r$ and $X_{LSA} = U_r S_r$ where $U_r, S_r, V_r$ are truncated matrices. In our case, this number will be chosen by cross validation. The intuition behind the use of LSA is that we want to discover groups of words that are equivalent in their meaning, as they should load onto the same singular vectors.

Even with the simple level of preprocessing provided by LSA, we can already start understanding how posters are separated into distinct topical areas. To visualize this, we transform the LSA vectors using a two dimensional reduction by $t$-Distributed Stochastic Neighbor Embedding ($t$-SNE) (\cite{van2008visualizing}, Fig. 1C), coloring each poster with its curated topical area from letter A through G. From the visualization, it is apparent that some topics are more clearly separable than others, at least as long as we are in a very low dimensional space.

\subsection{Poster representation based on keywords}

The dataset used here also contained keywords. Searching by keywords is a popular way to discover topics at conferences. We compare our methods to this type of search. In this algorithm, the LSA analysis is applied to the keyword vector of each poster and all the rest of the analysis described below is the same.

\subsection{Rocchio Algorithm and Nearest Neighbor Assignment}

The Rocchio algorithm is used to produce recommendations based on relevant and non-relevant documents previously voted by the user \cite{rocchio1971relevance}. The method can work with term frequency, tf-idf, or any term weighting schemes. Given a set of documents vectors that are found to be relevant by the user $\{r_1, r_2, \ldots, r_n\}$ and another set of non-relevant documents $\{u_1, u_2, \ldots, u_m\}$, the Rocchio algorithm finds a document vector that combines both sets of documents as follows

\begin{equation} \label{eq:1}
 q = q_0 + \frac{\alpha}{N} \sum_{i=1}^N r_i - \frac{\beta}{M} \sum_{j=1}^M u_j
\end{equation}
where $q_0$ was originaly proposed as the mean document vector of an initial query made into the system, such as a search for a particular keyword. The parameters $\alpha$ and $\beta$ control how much the relevant and non-relevant documents affect the final query vector, respectively. In the case of our set up, there is no such initial query $q_0$ and therefore the mean article replaces this term.

\begin{equation} \label{eq:2}
 q = \frac{\alpha}{N} \sum_{i=1}^N r_i - \frac{\beta}{M} \sum_{j=1}^M u_j
\end{equation}

The parameters $\alpha$ and $\beta$ will be found by cross validation using some external validation process.

Once the query vector $q$ has been defined for a user, a nearest neighbor search will be performed to retrieve suggested elements. In our implementation, we use ball trees to approximate a nearest neighbor search, which tradeoffs speed and accuracy relatively well \cite{shakhnarovish2005indyk}.

\subsection{Measuring suggestion performance by using human curated classification}

Each poster contained in this dataset has been manually classified into a topic by the organizers of the conference. The topic classification is based on a tree with three levels. For example, a poster might be classified with topic $F.01.r$, where $F$ is the broad area of study in Neuroscience, $01$ is a subarea of $F$, and $r$ is a further subdivision within $01$.  To validate many of the parameters in our performance tests, we use this topic classification as an indirect measure of how correct the suggestions are.

We implicitly assume that a random attendee would be highly interested in one topic only and not interested in others. This means that users would tend to like poster from the same area. While this assumption may not be accurate for a large number of attendees, we believe in captures the intention behind the classification of topics in this particular SfN conference. To measure the quality of suggestions, then, we ask the system to suggest ten posters based on a set of “liked” posters from a particular topic. For each of the suggestions, we compute the distance between its topic and the topic of the liked posters, using as distance measure the lowest common ancestor in the topic tree \cite{aho1976finding}. Minimizing the average of the humun topic distances between the liked poster and the ten suggested posters will be set as the performance metric for our comparisons.

\begin{algorithm}
\caption{\textbf{Science Concierge workflow with Latent Semantic Analysis}}
\begin{algorithmic}[1]
\State Receive set of $N$ documents $D = \{d_1, d_2, \cdots, d_N \}$
\State Preprocess documents $D = \mbox{preprocess}(\{d_1, d_2, \cdots, d_N \})$ using stemming and removing English stop words
\State Convert documents to matrix $X$ using term frequency, tf-idf or log-entropy. Remove rare words and words with high tf-idf
\State Apply Latent Semantic Analysis $X = USV^\top \rightarrow X \approx U_r S_r V_r$, $X_{LSA} = U_r S_r$
\State Index nearest neighbor model to $X_{LSA}$ or for word vector represention, use the average word vector of words in an abstract
\State Compute user's preference vector eq. (\ref{eq:2})
$$q = \frac{\alpha}{N} \sum_{i=1}^N r_i - \frac{\beta}{M} \sum_{j=1}^{M} u_j$$
\State Use indexed nearest neighbor model to suggest relevant documents
\label{algo:flow}
\end{algorithmic}
\end{algorithm}

\section{Results}

\subsection{Parameter optimization}

\subsubsection*{Number of components for LSA}

One parameter of the LSA algorithm is the number of components of the SVD decomposition \cite{bishop2006pattern}, which we find by cross validating on the distance to human curated topic classification as described in Methods. We randomly sampled one poster and liked it. Then, we ask the system to suggest ten posters and compute the average distance in human curated topic of those ten posters to the liked poster (Fig. 2). In this way, we were able to find that the appropriate number of components to be used should be around between 100 to 200. We select 150 components for later experiments described in paper.

\begin{figure}[!ht]
\centering
\includegraphics[width=3in]{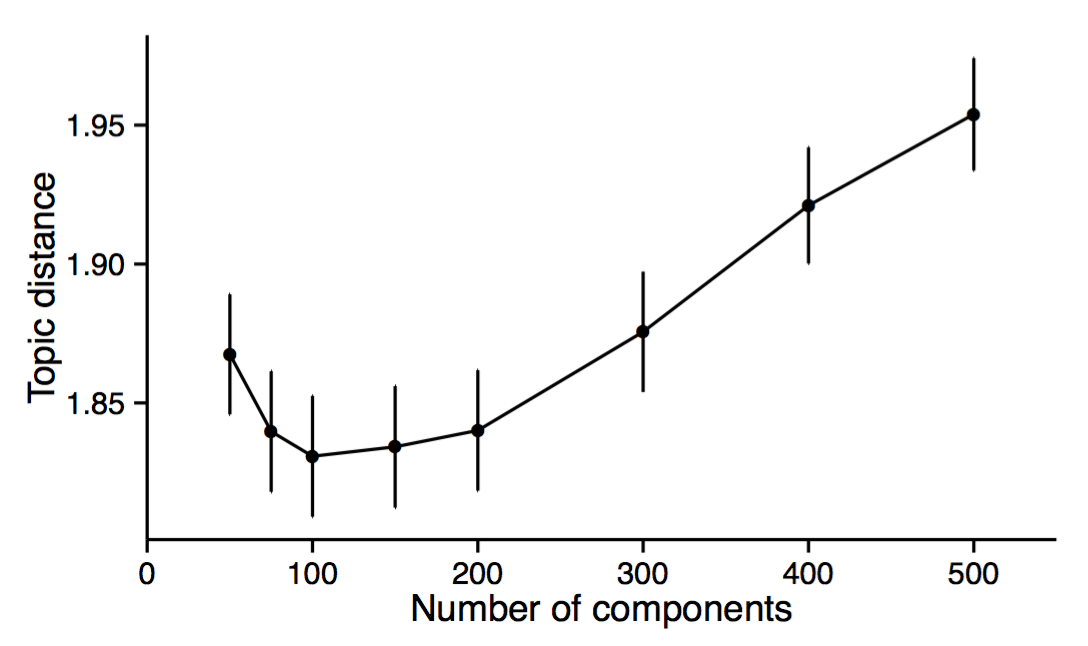}
\caption*{\textbf{Fig 2.} \textbf{Number of SVD components vs. performance of the algorithm to capture human curated topics}  The number of LSA components vs the average human curated tree distance of suggested posters.}
\end{figure}

\subsubsection*{Weight of relevant and non-relevant documents for Rocchio algorithm}

The Rocchio algorithm differently weighs the importance of relevant documents with a parameter $\alpha$ and the importance of non-relevant with a parameter $\beta$ (see  Eq. \ref{eq:1}). We tuned the parameters $\alpha$ and $\beta$ using a procedure similar to that used for finding the number of components in SVD.

The experiment this time is done by liking one poster from a random topic and voting as non-relevant a poster from a different topic. We performed three experiments where we tested three distances of the non-relevant voted posters: distance 1 (1 subdivision away), distance 2 (in a difference subarea), and distance 3 (in a different area of study). For each of these experiments, we tried a grid of $\alpha$ and $\beta$ parameters and computed the average topic distance over a set of 1,000 simulations (Fig. 3). This is possible because we only have two free parameters here.

We found that voting as non-relevant those posters that are 1 distance away produces large effects on performance. Interestingly, we found that the parameter $\beta$ almost always makes the performance decrease. This means that non-relevant posters should be used to filter out the recommendations rather modifying a user’s preference vector. We also found that the best values for the parameter $\alpha$ are always larger than 1. Additionally, we found that non-relevant posters that are 3 distance away in topic space have little effect on performance (Fig. 3). This is intuitive because disliking posters that are far away gives little information about the topic that a user may like. At the end, the best combination of parameters for non-relevant posters that are away 1 distance is $\alpha=1.8$ and $\beta=0$.

\begin{figure}[!ht]
\centering
\includegraphics[width=5in]{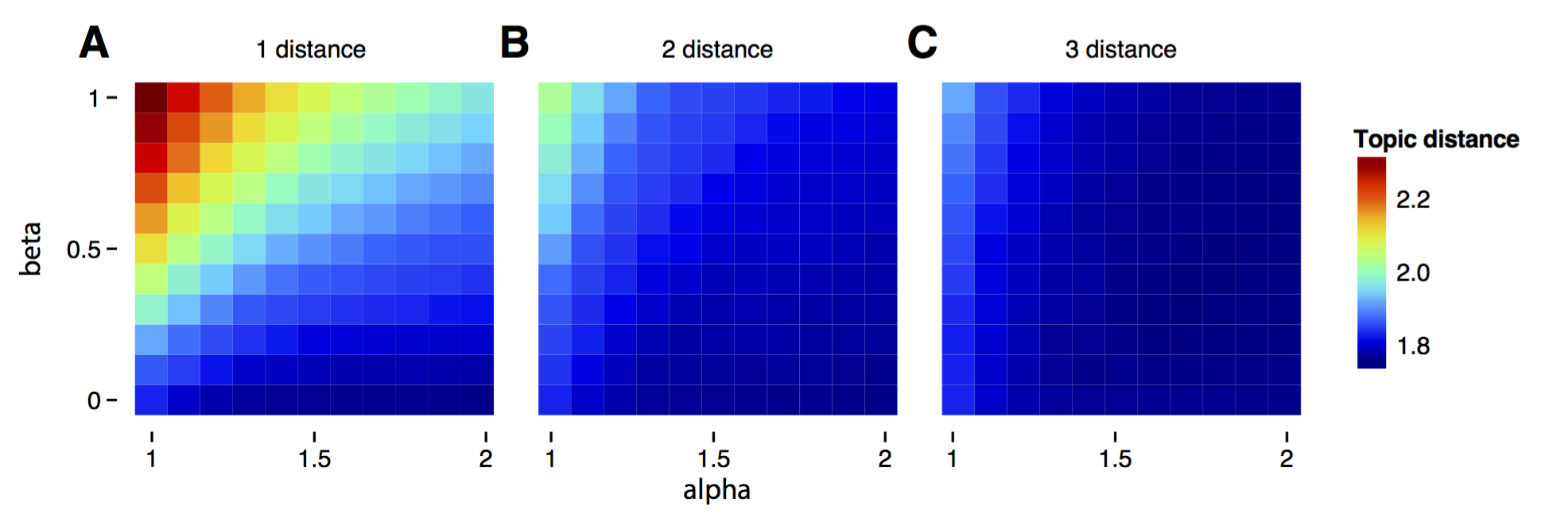}
\caption*{\textbf{Fig 3.} \textbf{Finding best parameters to weigh relevant and non-relevant votes}. Performance of the system as a function alpha and beta parameters for non-relevant documents that are 1 distance away (A), 2 dislike distance away (B), and 3 dislike distance away (C) in human curated topics.}
\end{figure}

\subsection{Algorithm comparison}

In this section, we compare the performance of our algorithm against common alternatives techniques and features. Our framework requires the use of abstracts, a significantly more complex data source than keywords, for example. Using LSA compromises speed in recommendation while still captures closeness in human curated distance after multiple votes. Moreover, using the full abstract could capture variability that is not available in other simpler methods. To properly test the advantages of our approach, we cross validate the performance of our method by using human curated topic distances as a performance metric (see section \ref{methods}).

For each of the algorithms tested below, we do the following simulation to estimate their performances. For each run of a simulation, we pick a poster at random and vote it as relevant. Then, we ask the algorithm to suggest ten posters based on that vote. We compute the average distance in human curated topic space between the suggestions and the liked poster. We then vote for another poster randomly selected from the same human curated topic as the first poster. We again ask the algorithm to suggest a set of ten posters and again we compute the average distance. We repeat this process ten times to obtain the average distance to human curated topics as a function of the number of votes. This simulation will help us understand the performance of the algorithms as they gather more votes from a simulated user.

As a baseline for all comparisons, we compute the performance of a null model that suggests posters at random. This is necessary because the distribution of human curated topics is not uniform and therefore the distances could be distributed in unexpected ways. Not surprisingly, the average distance to the human curated topic remained constant with the number of votes (Fig. 4, Random) but it was below 3, which is the farthest possible distance. This baseline will allow us to compare performance against a proper null model.

Keywords are a common recommendation technique that relies on using human curated “tags” for each document. In our dataset (see section \ref{methods}), the authors themselves assigned multiple keywords by picking them from a predefined set, crafted by the organizers. We used these keywords to produce  recommendations by treating them as documents. We then model keyword counts by transforming them from the tf-idf matrix to 30 SVD dimensions. Preliminary analysis revealed that 30 dimensions was an appropriate number to capture most of the variability. Suggestions using keywords perform significantly better than random suggestions (paired $t(9999) = 120.08$, $p < 0.0001$) and keyword suggestion performance improves with the number of votes, but not significantly (paired $t(9998) = -1.851$, $p = 0.064$, Fig. 4). While keywords allow for better suggestions than the null model, authors would need to manually provide them.

Science Concierge uses the abstracts to automatically extract topics and provide suggestions. We found that Science Concierge produces significantly better suggestions (Fig. 4) than keywords ($\Delta = 0.97$ $t(9999) = 138.26$, $p < 0.0001$),
term frequency
($\Delta = 0.3$ $t(9999) = 67.805$, $p < 0.0001$),
log entropy
($\Delta = 0.04$ $t(9999) = 13.987$, $p < 0.0001$),
and word vector
($\Delta = 0.4$ $t(9999) = 86.662$, $p < 0.0001$). For all techniques but
keywords, we found that the human topic distance of the recommendations
significantly decreases with the number of votes.

\begin{figure}[!ht]
\centering
\includegraphics[width=3.5in]{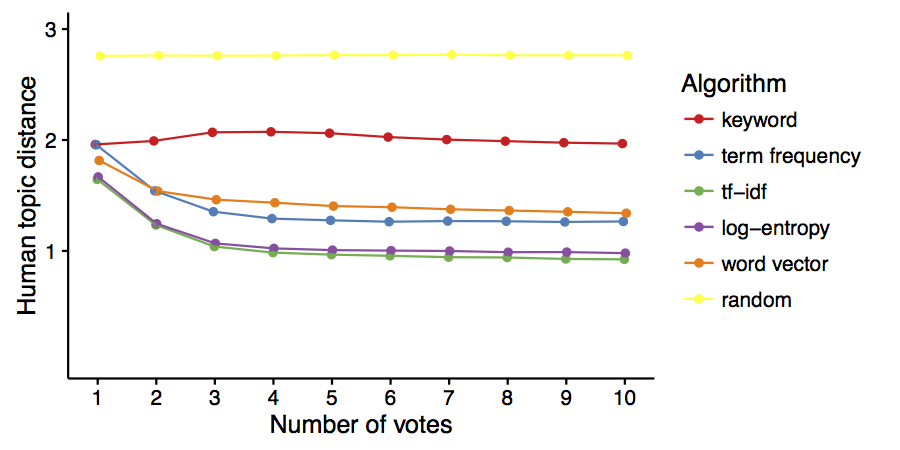}
\caption*{\textbf{Fig 4.} \textbf{Comparison of algorithms as they get more relevant documents from a simulated user}. All term weighting schemes except keywords
improve recommendations with votes}
\end{figure}

Ideally, we want distances between documents induced by an algorithm to closely match the distances induced by human curated topics. Here, we tested this idea using the keyword model and the Science Concierge model. This is, we tested the correlation between the distance of two random documents using a model and the distance of those same documents in human curated topic space.  Word vector had the highest correlation with human topics ($\rho = 0.48$), followed by tf-idf ($\rho = 0.43$), log-entropy ($\rho = 0.32$), term frequency ($\rho = 0.2$),
and keywords ($\rho = 0.17$). To compare these relationships on a similar scale, we visualized the human curated distance vs. the z-score of the model’s distances (Fig. 5).

\begin{figure}[!ht]
\centering
\includegraphics[width=3.5in]{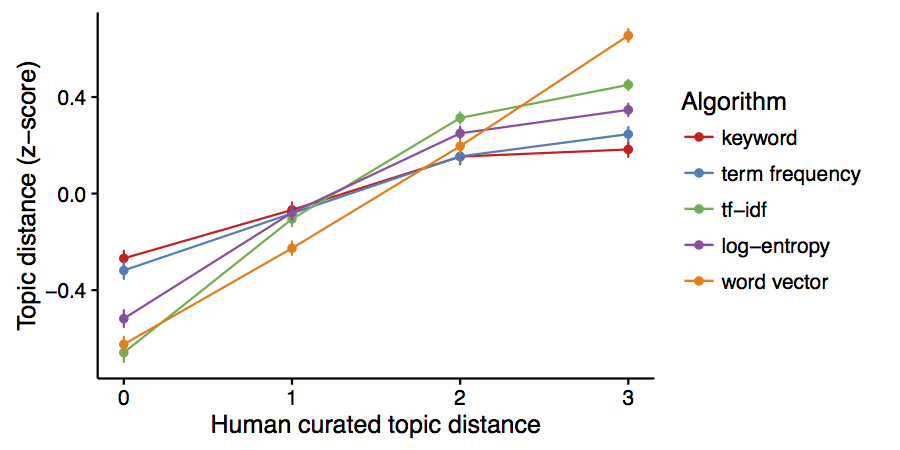}
\caption*{\textbf{Fig 5.} \textbf{Relationship between human curated distance and topic distance induced by the keyword and Science Concierge models}.
}
\end{figure}

\section{Discussion and conclusion}

Discovering new and relevant scholarly material progressively requires automation. Such discovery is already possible for
commercial content such as movies, news, and music. However, the same cannot be said about scientists. They commonly use legacy search systems that cannot learn from previous searchers. In this article, we propose a system that can improve recommendations based on a scientist votes for relevant and irrelevant documents. We tested the system on a set of posters presented at the Society for Neuroscience 2015 conference. We found that our system significantly improves on suggestions based on author-assigned keywords. We also found that our system learns as the user provides more votes. The system returns a complete schedule of posters to visit within 100 ms for a conference of around 15K posters. We publish the source code of the system so others can expand its functionality (\href{https://github.com/titipata/science_concierge}{\texttt{github.com/titipata/science\_concierge}}). In sum, this article presents a system that can make scientific discovery faster and it is openly available for other scientists to expand.

One surprising finding in our analysis is that the posters voted as non-relevant were better left not influencing the final preference vector. In particular, when voted non-relevant posters were close in topic space to the liked posters, then the performance degraded. If those non-relevant posters were far away, then the performance remains unchanged. In the future, we want to expand the experiment to domains in which a large number of votes is casted. This may allow the algorithm to better understand topic preferences and therefore offer suggestions that exploit the knowledge built in non-relevant votes.

The topic modeling technique used in our algorithm assumes that topics live on a linear space. While this assumption provides significant computational advantages, there might be cases where more complex modeling approaches may capture subtle topical relationships. In the future, we will try non-linear probabilistic approaches such as Latent Dirichlet Allocation (LDA), which recently has been shown to scale well \cite{hoffman2010online}. To better capture entities embedded in the text, future research will also investigate how to use deep learning modeling of words and phrases \cite{mikolov2013efficient, pennington2014glove}. However, it is unclear how these methods cope with scalability. Our system may already provide an appropriate level of speed and accuracy for our domain.

It may appear that the averaging of preferences that our
algorithm perform would mislead recommendations for scientists with
multiple, disparate interests. However, we found in the usage data
of the website that scientists tend to like few documents,
all within similar topics. More importantly, given the high
dimensional representation of the documents, averaging preferences
from different clusters would still produce sensible recommendations.
This happens because the nearest neighbor of the mean of two clusters is closer to those clusters than to a random document.
This phenomenon gets stronger with higher dimensions  \cite{hopcroftfoundations}. In our next research, we will explore more directly how to improve recommendations for scientists with diverse interests.

Future research could expand our system in many ways. The Rocchio algorithm’s dependency on nearest neighbors makes it inappropriate to exploit the potential richness available on large number of votes \cite{friedman2001elements}. With long term users who provide hundreds or even thousands of votes, it may be more accurate to cast the recommendations as a classification problem \cite{lavrenko2001relevance}. It is unclear however when would be the right time to make such as switch.

Our system proposes a new way to discover scholarly material. Many similar systems (e.g., Google Scholar) do not release their algorithms, making them difficult to study. By opening our algorithm, we will engage the scientific community to collaborate. Moreover, our algorithm does not necessarily need to be constrained to scientific text as any document that can be represented as a vector can be fed into it (e.g., scientific figures, audio). In sum, our system provides an open and fast approach to accurately discover new research.

\subsubsection*{Acknowledgements}

Titipat Achakulvisut was supported by the Royal Thai Government Scholarship grant \#50AC002. Daniel E. Acuna was supported by the John Templeton Foundation grant to the Metaknowledge Research Network. We thank SfN for providing the abstract dataset for our tests.

\subsubsection*{Author Contributions}

Titipat Achakulvisut (writing, programming, concept), Daniel E. Acuna (writing, programming, concept), Tulakan Ruangrong (programming), Konrad Kording (writing, concept).

\bibliography{citations}{}
\bibliographystyle{plain}

\end{document}